\documentclass[10pt,pra,amsmath,amssymb,aps,groupedaddress,notitlepage,twocolumn]{revtex4}

\usepackage{graphicx}
\usepackage{amssymb}
\usepackage{subfigure}
\usepackage{amsmath}
\usepackage{amsfonts}
\usepackage{bm}
\usepackage{color}
\usepackage{palatino}

\newcommand{\ket}[1]{\ensuremath{\left|{#1}\right\rangle}}

\graphicspath{{converted_graphics/}}
\begin{document}

\title{Experimental determination of multipartite entanglement with incomplete information}
\author{G. H. Aguilar}
\email{gabo@if.ufrj.br}
\affiliation{Instituto de F\' \i sica, Universidade Federal do Rio de Janeiro, Caixa Postal 68528, Rio de Janeiro, RJ 21945-970, Brazil}

\author{S. P. Walborn}
\email{swalborn@if.ufrj.br}
\affiliation{Instituto de F\' \i sica, Universidade Federal do Rio de Janeiro, Caixa Postal 68528, Rio de Janeiro, RJ 21945-970, Brazil}

\author{P. H. Souto Ribeiro}
\email{phsr@if.ufrj.br}
\affiliation{Instituto de F\' \i sica, Universidade Federal do Rio de Janeiro, Caixa Postal 68528, Rio de Janeiro, RJ 21945-970, Brazil}

\author{L. C. C\'{e}leri}
\email{lucas@chibebe.org}
\affiliation{Instituto de F\'{\i}sica, Universidade Federal de Goi\'{a}s, Goi\^{a}nia, GO, Brazil}


\begin{abstract}

Multipartite entanglement is very poorly understood despite all the theoretical and experimental advances of the last decades. Preparation, manipulation and identification of this resource is crucial for both practical and fundamental reasons. However, the difficulty in the practical manipulation and the complexity of the data generated by measurements on these systems increase rapidly with the number of parties. Therefore, we would like to experimentally address the problem of how much information about multipartite entanglement we can access with incomplete measurements. In particular, it was shown that some types of pure multipartite entangled states can be witnessed without measuring the correlations [M. Walter \textit{et al.}, Science \textbf{340}, 1205 (2013)] between parties, which is strongly demanding experimentally. We explore this method using an optical setup that permits the preparation and the complete tomographic reconstruction of many inequivalent classes of three- and four-partite entangled states, and compare complete versus incomplete information. We show that the method is useful in practice, even for non-pure states or non ideal measurement conditions.  
\end{abstract}

\maketitle

\section{Introduction}

According to quantum mechanics, the state of a system can be represented by a linear combination of different eigenstates of an observable. This fact, known as the superposition principle, prevents us from constructing a representation of physical reality based on our classical intuition. When applied to composite systems, this principle leads to the fundamental concept of entanglement. Essentially, when quantum objects interact they can no longer be described by individual independent states. Rather, they are instead a superposition of tensor product states.  In other words, entangled parties cannot be treated as independent systems with well defined physical properties \cite{Horodeki}.

On a fundamental level, entanglement is a geometric consequence of the replacement of the classical phase space by the quantum projective Hilbert space, presenting a richer structure whose complexity grows exponentially with the number of parties \cite{Horodeki}. Even for the simplest case of a bipartite system, in which well-defined measures of entanglement and its relation with information processing tasks are well understood, the theory still exhibits some puzzles, such as the phenomenon of entanglement locking \cite{Horodecki1}. For the multipartite case, several new difficulties arise. For instance, many inequivalent classes of entanglement are possible \cite{Bennett1,durr00,verstraete01}. How to theoretically identify and experimentally distinguish such classes is one of the fundamental problems in this field. Another issue is related to the fact that the number of measurements, measurement time, and computational effort for processing the tomographic data of a multipartite state scales exponentially with the number of qubits. 

The present work contributes to the understanding of these problems considering three points. First, we provide a practical photonic scheme, based on the entanglement between two different degrees of freedom of photon pairs to prepare and measure genuinely entangled states of three and four qubits in a controlled way. This setup allows us to compare the local and the global information obtained from the same set of measurements. Second, we test the limits of validity of a witness for multipartite entanglement \cite{Walter} in a real laboratory scenario, in contrast to the ideal case of pure states. We test this requirement experimentally and show that in our data the criteria remains useful to study the properties of different kinds of entanglement, even under moderate noise. In practice, witnessing multipartite entanglement and being able to tell the class of a given state might find direct application to quantum information protocols that require specific types of entanglement. Third, we observe that this approach outperforms experimental methods that obtain complete tomographic information of the quantum state. It is important to note here that we still need an upper bound on the purity of the global state in order to be confident about the witness, but this still requires less resources than the full quantum state tomography (see Ref. \cite{Walter} and Appendix C for more information). This improvement is not only related to the reduced number of measurements, but also to the reduced sensitivity to imperfections like non unity detection efficiency.

\section{Theory}

Motivation for this approach arises from a typical scenario in quantum information processing, when several parties share a global quantum state and they are allowed to locally act on each individual system and to communicate classically (local operations and classical communications --- LOCC). Among several actions, they could wish to transform the total entangled state into another. This kind of situation leads to natural ways of defining distinct equivalence classes of entanglement. Considering only a single copy, two pure quantum states can be obtained from each other through LOCC only if they are related by local unitaries, which leads to an infinite amount of equivalence classes of entanglement, even for the bipartite case (we need continuous parameters to label all the classes) \cite{Linden}. For instance, one pure non-maximally entangled state can be converted into another state with the same amount of entanglement using LOCC (if the entanglement decreases, which is possible with LOCC, we cannot revert the operation and both states would not belong to the same class), and this defines one class of entanglement. Because the coefficients of a non-maximally entangled state are continuous parameters, we have an infinite amount of states, each one defining an equivalence class of entanglement. A coarse-grained classification defines that two states are equivalent ---in the sense that they posses the same kind of entanglement--- if they can be converted to each other by LOCC with a \emph{finite} probability of success \cite{Bennett1,durr00}. Mathematically, two pure quantum states $|\phi\rangle$ and $|\psi\rangle$ are equivalent if and only if there exists invertible local operators $\lbrace O_{i}\rbrace_{i=1}^{N}$ such that $|\phi\rangle = O_{1}\otimes\cdot\cdot\cdot\otimes O_{N}|\psi\rangle$ \cite{durr00}. These operators are the so-called stochastic LOCC, or SLOCC \cite{Bennett2}, and lead to a finite classification of multipartite systems into distinct \emph{families} of entanglement. It is also convenient from the experimental point of view due to the fact that states belonging to the same class are suitable for performing the same task (although the probability of success may differ). In Ref. \cite{Walter} a new classification scheme based on the geometry of the eigenvalue space and local measurements on the subsystems was proposed. This new classification defines different classes of entanglement, always being finite.

The scheme introduced by Walter \emph{et al.} \cite{Walter} is based on the solution of the quantum marginal problem \cite{Schilling,Schilling1}. For instance, let us consider a multipartite state $\rho$ describing the state of $N$ qubits. One can ask which set of single-party density matrices are compatible with $\rho$. By compatible we mean that there exist reduced one-party density matrices $\rho_{i}$ such that $\rho_{i} = \mbox{Tr}_{\bar{i}}(\rho)$, where $\mbox{Tr}_{\bar{i}}$ is the trace over all but the $i$-th part. This is known as the quantum marginal problem \cite{Schilling}, which has been completely solved in Ref. \cite{Higuchi} for the case of $N$ qubits, given that the global state is pure. The solution of this problem in the general case is practically intractable \cite{Liu}. For the case of an $N$-qubit pure state considered here, the spectrum of the reduced density matrices must satisfy the so-called polygon inequalities \cite{Higuchi}
\begin{equation}
\lambda_{k} \geq \sum_{\substack{i = 1\\ i \neq k}}^{N}\lambda_{i} - (N - 2),
\label{polygon}
\end{equation}
where $\lambda_{k} \in [1/2,1]$ is the \emph{maximum} eigenvalue of $\rho_{k}$. Note that these inequalities determine the complete set of all possible reduced one-party density matrices since the maximum eigenvalue completely characterizes the reduced $2\times 2$ density matrix of each qubit (assuming normalization). 

It was recognized that all possible sets of solutions of the quantum marginal problem
\begin{equation}
\Lambda = \left(\lambda_{1},...,\lambda_{N}\right)
\end{equation}
form a convex polytope \cite{Christandl,Klyachko}. For the cases where the global state is (almost) pure, these local eigenvalues contain considerable information about the entanglement of $\rho$. The set of possible $\Lambda$'s associated with global states restricted to a given entanglement class also forms a convex polytope, the so called \emph{entanglement polytope}. Therefore, if $\Lambda_{\rho}$ does not belong to a given entanglement polytope $\Delta_{\mathcal{C}}$, then $\rho$ cannot belong to the associated entanglement class $\mathcal{C}$:
\begin{equation}
\Lambda_{\rho} \not\in \Delta_{\mathcal{C}} \Rightarrow \rho \not\in \mathcal{C}.
\label{witness}
\end{equation}
Note that it is not always possible to determine two states as belonging to two inequivalent classes. This is due to the fact that the entanglement classes forms a natural hierarchy, with one class contained inside the other, as discussed in more detail below (see also the Appendix). It is important to state here that we still have to deal with infinitely many entanglement classes (except for the simple cases of $N \leq 3$). However, the result of Ref. \cite{Walter} shows that, despite this fact, we always have a finite number of entanglement polytopes. Exploiting the convexity of these polytopes, the authors of Ref. \cite{Walter} arrived to a witness criteria (see Appendices for more details).

We analyse three and four qubits cases, and show that identification of genuine entanglement is possible even in the presence of moderate noise. We will discuss the role of the purity of the states in this method from the perspective of our experimental realization below. For clarity, in the main part of the text we describe the experimental procedure and the results for three qubits, leaving the four qubit case, which is longer though analogous, to the Appendix. 

\subsection*{The polytopes for three qubits}

For three qubits we have six different entanglement classes, two of them containing genuine three-partite entanglement \cite{durr00} (see also the Appendix). Recalling that we are dealing with initial pure states, these classes are defined as follows. 
\begin{enumerate}

\item \emph{The fully separable states} ($\mathcal{S}$) --- These states can be represented, in the computational basis, by $|\psi\rangle_{\mathcal{S}} = |000\rangle$. Therefore, any fully separable state can be converted into $|\psi\rangle_{\mathcal{S}}$ by means of a convenient SLOCC protocol. In this case, all the reduced states are also pure, and we must have
\begin{equation}
\Lambda_{\mathcal{S}} = (1,1,1).
\end{equation}
That is, the entanglement polytope is a single point, the upper vertex of the tetrahedron of Fig. \ref{poly3}(a). All the reduced local density operators are represented by rank one matrices.

\begin{figure*}[tbp]
\includegraphics[width=1\textwidth]{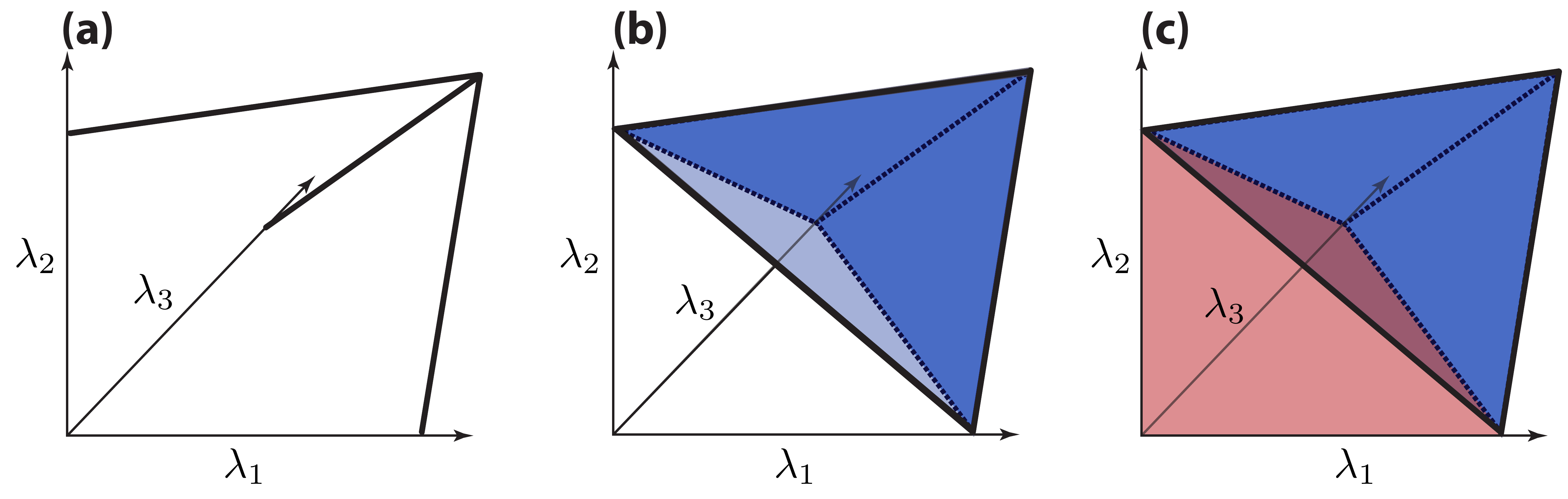}
\caption{Entanglement polytopes for the three qubit case. a) Polytope for the biseparable and separable cases ($\mathcal{BS}$  and $\mathcal{S}$ states). There are three axis for the eigenvalues $\lambda_1$, $\lambda_2$ and $\lambda_3$, and the other three lines represent the $\mathcal{BS}$ polytopes. Each line corresponds to one polytope. These lines converge to a point, which represents the polytope of the full separable states. b) Polytope for the $\cal{W}$ states, represented by the blue tetrahedron, a 3-D representation. c) Polytopes for the $\cal{W}$ states (the blue tetrahedron) and the polytope for the GHZ states (the entire polytope). Points inside the red tetrahedron are guaranteed to belong to the $\mathcal{GHZ}$ class.}
\label{poly3}
\end{figure*}

\item \emph{Bi-separable states} ($\mathcal{BS}$) --- Apart from permutations of the parties and local unitaries, these three classes can be represented, for instance, by the state $|\psi\rangle_{\mathcal{BS}} = |0\rangle \left(\alpha|00\rangle + \beta|11\rangle\right)$, with $|\alpha|^{2} + |\beta|^{2} = 1$. Considering only one of these cases (the other two are obtained by simple permutation of the labels), the possible set of eigenvalues are then given by
\begin{equation}
\Lambda_{BS} = (1,\lambda_{2},\lambda_{3}),
\end{equation}
with $\lambda_{2},\lambda_{3} \in [0.5,1)$. This leads us to the entanglement polytope defined by $2 \leq 1 + \lambda_{2} + \lambda_{3} \leq 3$. Moreover, in order to satisfy Eqs. (\ref{polygon}) we must have $\lambda_{2} = \lambda_{3}$. These entanglement polytopes are represented by the thick straight lines of Fig. 1(a), the ones originating in the upper vertex. In other words, if one finds that one of the local measured eigenvalues is equal to one, by the witness criteria (\ref{witness}) the global state does not present genuine multipartite entanglement ---whether the state is fully separable or it is bi-separable. From the above relations we can see that if all the local eigenvalues are smaller than one, we must have genuine multipartite entanglement, which is divided into two inequivalent classes. 

\item \emph{the $\mathcal{W}$ states} --- can be represented by  
\begin{equation}
|\psi\rangle_{\mathcal{W}} = \frac{1}{\sqrt{3}}(|001\rangle + |010\rangle + |100\rangle),
\label{Wstate}
\end{equation}
and the associated polytope is determined by the relation
\begin{equation}
\lambda_{1} + \lambda_{2} + \lambda_{3} \geq 2,
\end{equation}
together with Eqs. (\ref{polygon}), and is shown in Fig. \ref{poly3}(b) (the blue tetrahedron).

\item \emph{the $\mathcal{GHZ}$ states} --- can be represented by 
\begin{equation}
|\psi\rangle_{\mathcal{GHZ}} = \frac{1}{\sqrt{2}}(|000\rangle + |111\rangle).
\label{GHZ_state}
\end{equation}
The $\mathcal{GHZ}$ polytope is the entire polytope (blue plus red tetrahedrons in Fig. \ref{poly3}(c). However, accordingly to the definition of entanglement polytopes, if, for a given state, its local maximum eigenvalues respect the relation
\begin{equation}
\lambda_{1} + \lambda_{2} + \lambda_{3} < 2,
\label{GHZ_polytope}
\end{equation} 
together with the constraints imposed by Eq. (\ref{polygon}), we are sure that this state belongs to the $\mathcal{GHZ}$ class. This inequality determines the polytope illustrated in Fig. \ref{poly3}(c) (the red tetrahedron). Thus to determine that a given state contains indeed $\mathcal{GHZ}$-type entanglement, the experimental point must not be located inside any other polytope.
\end{enumerate}

It is a mathematical consequence of the definition of the equivalence classes of entanglement adopted in Ref. \cite{Walter} that these entanglement polytopes respect a natural hierarchy.
\begin{equation}
\Delta_{\mathcal{S}} \subseteq \Delta_{\mathcal{B}\mathcal{S}} \subseteq \Delta_{\mathcal{W}} \subseteq \Delta_{\mathcal{G}\mathcal{H}\mathcal{Z}}. 
\end{equation}
Geometrically, the relation $\Delta_{\mathcal{X}} \subseteq \Delta_{\mathcal{Y}}$ tells us that states from class $\mathcal{X}$ can be arbitrarily approximated by states in the class $\mathcal{Y}$ using SLOCC. This can be seen in Fig. \ref{poly3}. If the experimental set of eigenvalues lies, for instance, inside blue tetrahedron, we cannot conclude whether the associated global state belongs to the $\mathcal{W}$ or to the $\mathcal{GHZ}$ class, but we know that it contains genuine three-partite entanglement. However, if the experimental point is located inside the lower tetrahedron of \ref{poly3}\textbf{(c)} we can safely say that the corresponding state must contain genuine $\mathcal{GHZ}$ entanglement. 

\section{Experiment and Results}

\subsection{Experimental setup}

\begin{figure}[tbp]
\centering 
\includegraphics[width=8.7cm]{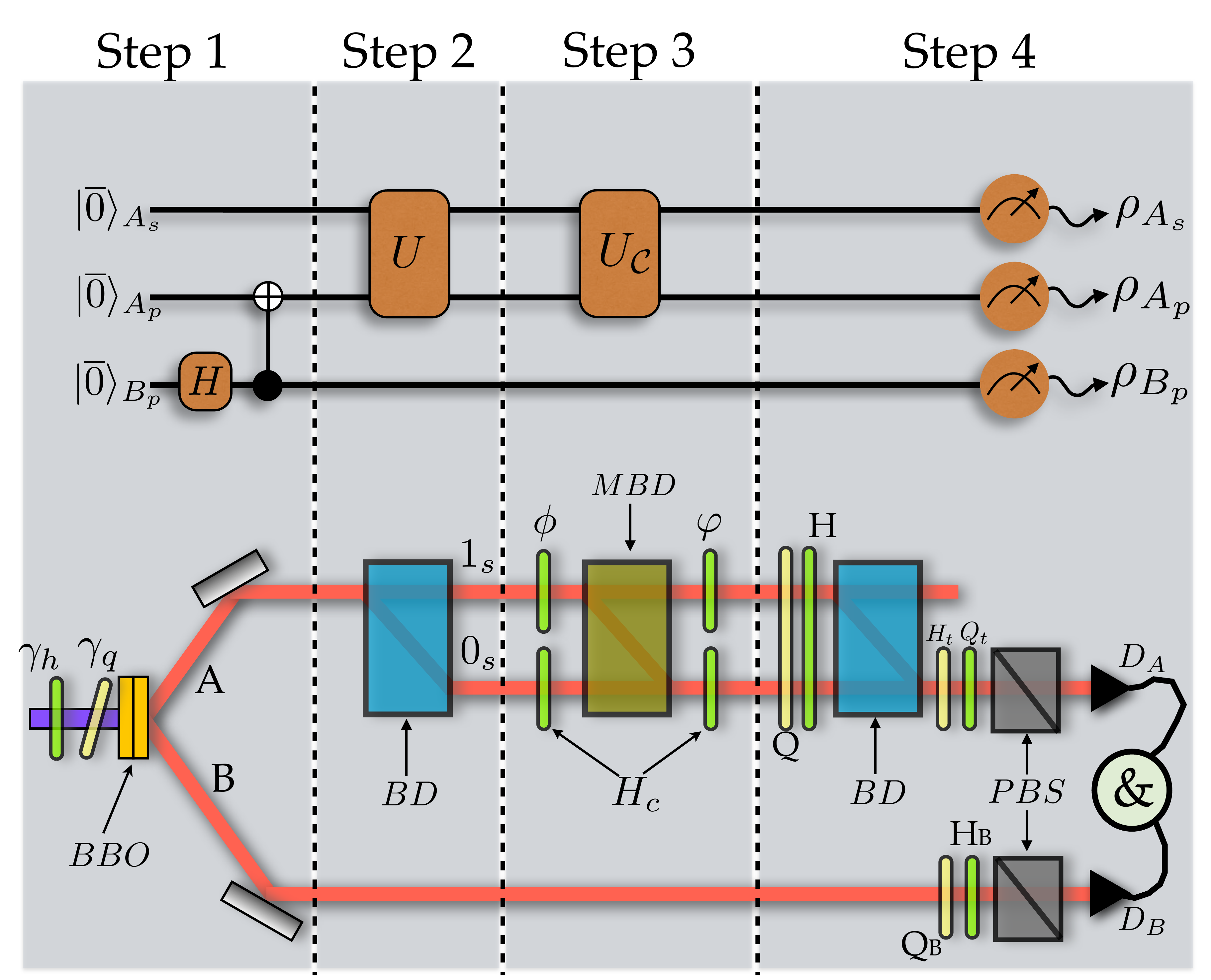}
\caption{Experimental Setup. The top panel shows the quantum circuit employed in our experiment. In the main text we give a complete description of each step in this circuit as well as its implementation in the optical scenario, shown in the bottom panel of the figure. The parameters $\alpha$ and $\beta$ appearing in Eq. (\ref{initial}) can be controlled by the half-wave ($\gamma_{h}$) and quarter-wave ($\gamma_{q}$) plates located before the $BBO$ crystals. The $BD$s are beam displacers and the $PBS$s are polarizing beam splitters. The modified beam-displacer ($MBD$), in contrast with the ordinary $BD$, transmits the horizontal polarization and deflects the vertical one. Physically, it is implemented by placing a $BD$ between two half-wave plates, which are not shown in the figure. The $\phi$ and $\varphi$ plates are used to ensure identical path lengths in the interferometers. $H_{i}$ and $Q_{i}$ are half- and quarter-wave plates used in the tomographic process, as explained in the text, and the symbol $\&$ represents coincidence counting.}
\label{fig:setup}
\end{figure}

The experimental setup is similar to the one used in the investigation of entanglement dynamics in Refs. \cite{farias12b, aguilarPRA14}. First, we will be restricted to the creation and measurement of different inequivalent classes of three-qubit states. The case of four qubits follows a similar procedure and it is presented in the Appendix. The main idea is to use twin photons, which are entangled in the polarization degree of freedom (represented by a subindex $p$ in the text), and to perform some operations to entangle this degree of freedom with the spatial mode (represented by the subscript $s$ in the text) of one or both photons, to produce three or four-partite entangled states respectively. 

A simplified scenario can be seen in the circuit diagram on the top of Fig \ref{fig:setup}. We begin with all the qubits initialized in the state $\ket{\bar{0}}_{i}$, with $i = \text{A}_p, \text{B}_p, \text{A}_s$ (this represents a general initial state and not necessarily the usual computational basis employed throughout the text). In Step 1 we implement a Hadamard ($H$) and a $CNOT$ gate in qubits $A_p$, $B_p$ producing a global state that is entangled in the $A_pB_p$ partition and separable with respect to $A_s$. In Step 2, a unitary operation $U$ is applied to qubits $A_pA_s$ with the purpose of creating entanglement between all the three qubits. In Step 3, we modify the entanglement class by applying the unitary transformation $U_\mathcal{C}$ on qubits $A_p$ and $A_s$. At the output, depending on the parameters defining $U_\mathcal{C}$, we can create states of all the inequivalent classes of three qubits \cite{durr00}. Step 4 is the measurement step. We perform quantum state tomography of the local states, thus reconstructing the individual reduced matrices $\rho_j$ (incomplete information), or of the the global state $\rho$ (complete information).

The implementation of this quantum circuit in the optical system is depicted in the bottom of Fig. \ref{fig:setup}. The four steps are as follows:

\emph{Step 1} --- With a 325 nm laser, we pump two crossed-axis Type I $\beta$-Barium Borate ($BBO$) crystals and create photons in a state close to \cite{kwiat99}:
\begin{equation}
\ket{\Psi}=(\alpha\ket{0}_{A_p}\ket{0}_{B_p}+\beta\ket{1}_{A_p}\ket{1}_{B_p})\ket{0_{s}}_{A_s}
\label{initial}
\end{equation}
where $\ket{0}_{B_p}$ ($\ket{1}_{B_p}$) is the horizontal (vertical) polarization of photon $B$. $\ket{0_{s}}_{A_s}$ represents the initial state of the spatial degree of freedom of photon $A$. The values of the amplitudes $\alpha$ and $\beta$ can be controlled by manipulating the polarization of the pump laser. This was done using a half-wave plate ($HWP$) $\gamma_{h}$ and the quarter-wave plate ($QWP$) $\gamma_{q}$ (see Fig. \ref{fig:setup}). Photon $A$ is directed to a nested interferometer which implements all operations described in the circuit diagram. Photon $B$ is sent to a polarization analysis which happens at Step 4.

\emph{Step 2}---   The first unitary operation, $U$, is applied by the Beam-Displacer $BD$, which implements the transformations $\ket{0}_{A_p} \ket{0_{s}}_{A_s}\rightarrow\ket{0}_{A_p} \ket{0_{s}}_{A_s}$ and $\ket{1}_{A} \ket{0_{s}}_{B}\rightarrow\ket{1}_{A_p} \ket{1_{s}}_{A_s}$. After this operation, the global state can be written as
\begin{equation}
\ket{\Psi}=\alpha\ket{0}_{A_p}\ket{0}_{B_p}\ket{0_s}_{A_s} + \beta\ket{1}_{A_p}\ket{1}_{B_p}\ket{1_s}_{A_s},
\label{GHZ}
\end{equation} 
which, by choosing $\alpha=\beta=1/\sqrt{2}$, is the $\mathcal{GHZ}$ state of Eq. (\ref{GHZ_state}), genuinely entangled in all the qubits \cite{ghz93}. 

\emph{Step 3} --- A unitary operation $U_{\mathcal{C}}$ is implemented by a set of $HWP$  and a Modified-Beam-Displacer $MBD$ (see the bottom panel of Fig. \ref{fig:setup}). The complete transformations can be written as
\begin{eqnarray}
\ket{0}_{A_p} \ket{0_{s}}_{A_s} & \rightarrow & \ket{0}_{A_p} \ket{0_{s}}_{A_s}  \nonumber \\
\ket{1}_{A_p} \ket{1_{s}}_{A_s} & \rightarrow & \cos{2\phi}\ket{1}_{A_p} \ket{0_{s}}_{A_s}  \nonumber \\ 
&-&\sin{2\phi}(\cos{2 \varphi} \ket{0}_{A_p} \ket{1_{s}}_{A_s} \\ 
& - &  \sin{2 \varphi} \ket{1}_{A_p} \ket{1_{s}}_{A_s})\nonumber,
\label{eq:unitary}
\end{eqnarray}
where $\phi$ and $\varphi$ are the rotation angles of the $HWP$'s shown in Fig. \ref{fig:setup}. Both $HWP$ and $H_c$ do not introduce rotations in the polarization and are used to compensate the optical length of the different paths. The global state of the system can be written as
\begin{equation}
\begin{split}
&\ket{\Psi}_{A_pB_pA_s}=\alpha\ket{000_s}+\beta[ \cos(2\phi)\ket{110_s} \\ 
& -\sin(2\phi)(\cos(2 \varphi) \ket{011_s} - \sin(2 \varphi) \ket{111_s})].
\end{split}
\label{constructed_state}
\end{equation}
We can see from Eq. (\ref{constructed_state}) that, by choosing different values in the set $\mathcal{M}=[\gamma_{h},\gamma_{q},\varphi, \phi]$, we are able to construct  states of three qubits with different types of entanglement. For instance, two different bi-separable states are obtained when $\mathcal{M}=[\pi/2,0,0,0]$ and $\mathcal{M}=[\pi/2,0,\pi/4,0]$. By choosing $\mathcal{M}=[\pi/3,0,\pi/8,0]$ we obtain the state $1/\sqrt{3}(\ket{000_s}+\ket{110_s}+\ket{011_s})$, which corresponds to the state of Eq. (\ref{Wstate}) with the second qubit flipped. To obtain states similar to the one in Eq. (\ref{Wstate}), we simply apply a rotation on qubit $B_p$ using a $HWP$ (not shown). Finally, a state that corresponds to the $\mathcal{GHZ}$ class is obtained when $\mathcal{M}=[\pi/2,0,\pi/4,\pi/4]$. 

\emph{Step 4} ---  After the engineering of distinct entanglement classes, we make projective measurements in the different degrees of freedom. The $BD$ has two  important tasks. On the one hand, together with the $H$ and $Q$, it is used to make projective measurements on the polarization degree of freedom. In this sense, the $BD$ is used as a polarizer. On the other hand, the $BD$ coherently combines the spatial modes $\ket{0_{s}}$ and $\ket{1_{s}}$ \cite{farias12b, aguilarPRA14}. Using this, we were able to perform complete tomographic measurements of the whole tripartite system as well as of the states of the individual systems. After the projection in the polarization and spatial mode, the photons are detected in $D_A$ and $D_B$ and coincidence counts are registered. In this way, states belonging to the six different classes of entanglement can be measured.

\subsection{Experimental results}

\begin{figure}
\includegraphics[width=8cm]{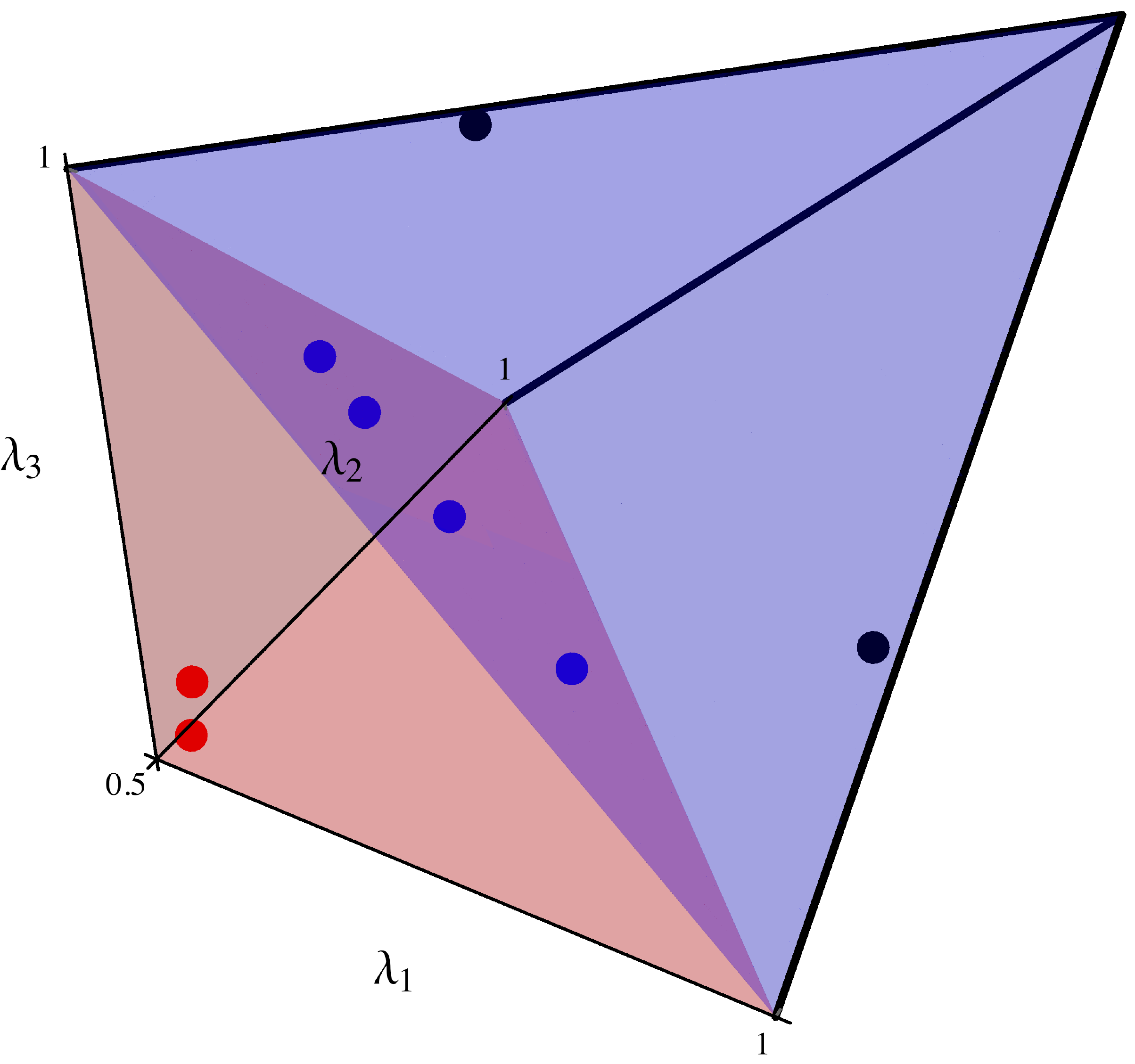}
\caption{Full polytope for three qubits. The blue region is the $\mathcal{W}$ polytope while the light red one represents all the $\mathcal{GHZ}$ states that cannot contain $\mathcal{W}$ kind of entanglement (see also Fig. \ref{poly3}). The blue red and black symbols represent, respectively, the $\mathcal{W}$, the $\mathcal{GHZ}$ and the $\mathcal{BS}$ states. The error bars are inside each point (see Supplementary Material for details).}
\label{FigWit}
\end{figure}

We performed quantum state tomography on every local qubit for different configurations of $\mathcal{M}$. The local density matrices were reconstructed using the maximum likelihood method, and the largest eigenvalues were discovered \cite{james01}. The results of this procedure are shown in Fig. \ref{FigWit}. As described before, the entire shaded region represents the $\mathcal{GHZ}$ polytope, the black lines represent the polytopes of the bi-separable states, and the upper blue region corresponds to the $\mathcal{W}$ polytope. The red shaded area corresponds to the region where only the $GHZ$ entanglement class can be found. In addition, we performed full quantum state tomography of each state for comparison, obtaining purities higher than $0.87$ in all cases (see the Appendix for the purity of all the prepared states).  
\par
By choosing $\mathcal{M}=[\pi/2,0,0,0]$ and $\mathcal{M}=[\pi/2,0,\pi/4,0]$ we create two different bi-separable classes experimentally. By tomographic reconstruction of the global state, we can confirm that these states are indeed bi-separable (by comparing the reconstructed state with the theoretical prediction). The local eigenvalues were obtained from the local density matrices alone, and are represented by the black dots in Fig. \ref{FigWit}. We can see that these states are close to the lines corresponding to the bi-separable polytopes. 
\par
We also produced different states of the $\mathcal{W}$ entanglement class. This was done by choosing four distinct values of the set $\mathcal{M}=[\pi/3,0,\varphi,0]$. Using the full density matrices obtained from full quantum state tomography, we calculate fidelities higher than 0.87 with respect to the $\mathcal{W}$ states given in Eq. \eqref{Wstate}.  This is sufficient to confirm the presence of multipartite entanglement of the $\mathcal{W}$ class \cite{acin01}. The experimental results of the local eiqenvalues are plotted with blue dots in Fig. \ref{FigWit}. We can see that these points are in the lower limiting area of the $\mathcal{W}$ polytope. For these states we can guarantee the presence of genuine multipartite entanglement. However, we cannot confirm whether it is of the $\mathcal{W}$ or $\mathcal{GHZ}$ type. This is a property of the witness \eqref{witness} and not caused by experimental imperfections. 
\par
By selecting $\mathcal{M}=[\pi/2,0,\pi/4,\pi/4]$ we measure states in the $\mathcal{GHZ}$ class. The results obtained from the local tomography are plotted with red dots. Note that both points are located in the light red area assuring the presence of multipartite entanglement of the $\mathcal{GHZ}$ class. Note that the these points are close to the $(\lambda_1,\lambda_2,\lambda_3)=(1/2,1/2,1/2)$, which corresponds to the pure $\mathcal{GHZ}$ state. To confirm this local information, we reconstructed the global density matrix obtaining a fidelity with respect to the  $\mathcal{GHZ}$ state higher than 0.85, proving the presence of genuine $\mathcal{GHZ}$ entanglement \cite{acin01}.
 
\begin{figure}[h]
\includegraphics[scale=0.4]{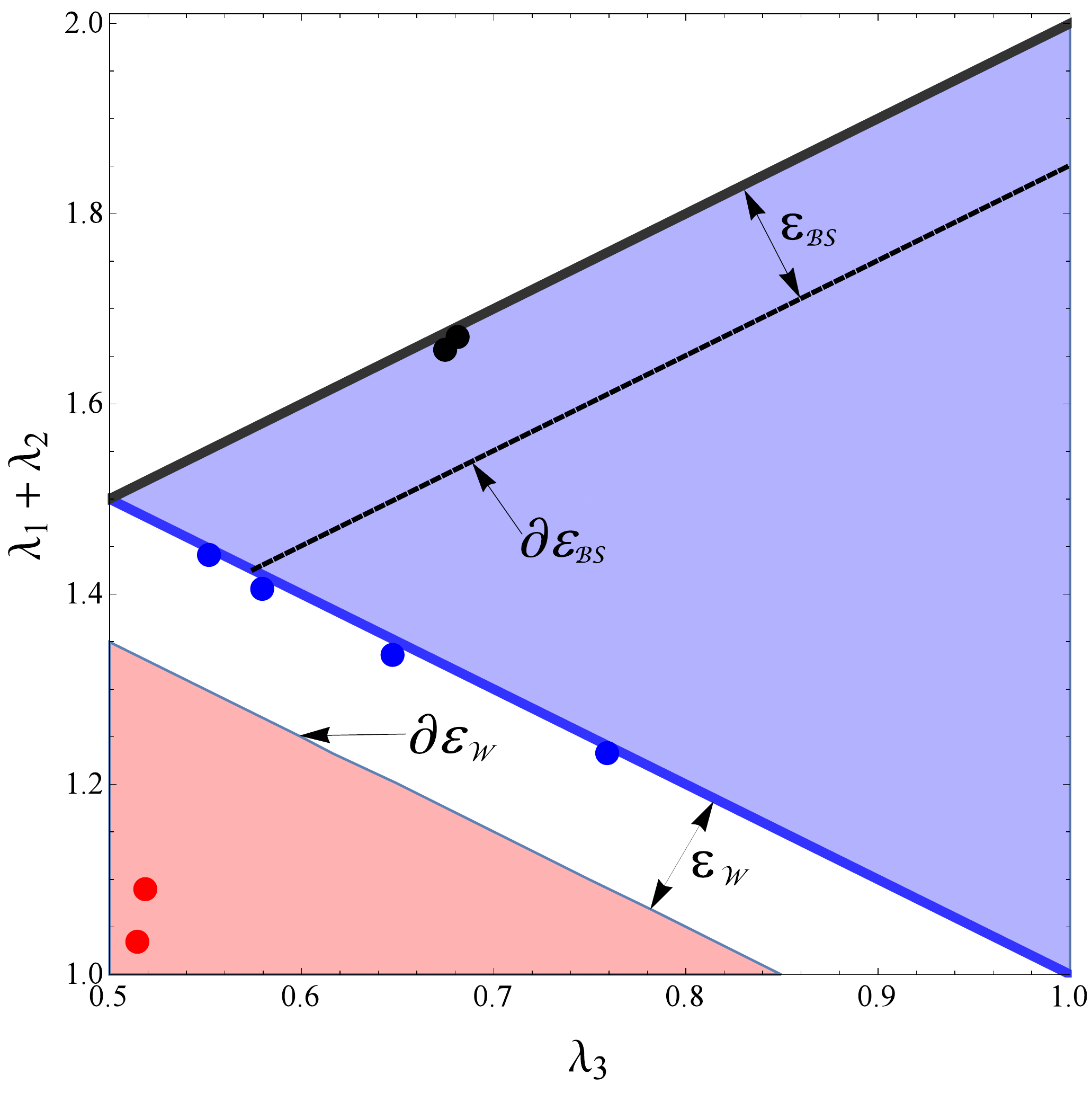}
\caption{Projected eigenvalue space --- The red region represents the $\mathcal{GHZ}$ polytope, and the blue region the $\mathcal{W}$ polytope. The black line represents the $\mathcal{BS}$ polytopes. The vertical axis contains a sum of two eigenvalues. For the case of the $\mathcal{BS}$ states it always contain the maximum eigenvalue, which is close to one, in such a way that all the three classes of $\mathcal{BS}$ states are projected onto the same line. As in Fig. \ref{FigWit}, the blue dots are the $\mathcal{W}$ states while the red and black ones represent the $\mathcal{GHZ}$ and the $\mathcal{B}\mathcal{S}$ states, respectively. The white region between the $\mathcal{GHZ}$ and $\mathcal{W}$ polytopes (whose theoretical border $\partial\mathcal{W}$ was obtained considering pure states) is the \emph{error} region $\varepsilon_{\mathcal{W}}$, which is a function of the purity of the experimental density matrix (see the Appendix for details). The border for mixed states is represented by $\partial\varepsilon_{\mathcal{W}}$. The same pattern is applied to the border between the $\mathcal{B}S$ and $\mathcal{W}$ polytopes (the dashed line in the figure). To compute the size of this region we chose the lowest purity among all the prepared states.}
\label{FigErr}
\end{figure}

As mentioned above, the witness analysed in Fig. \ref{FigWit} presupposes pure global states, which is a very strong assumption. However, it was shown \cite{Walter} that this witness should be robust against low levels of noise. Let us describe now how the experimental errors were taken into account, and analyse  the confidence of the witness with respect to the purity of the experimental states. In the presence of noise, the entanglement polytopes are transformed in a simple manner.  For instance, Eq. (\ref{GHZ_polytope}) changes to $\lambda_1(\rho)+\lambda_2(\rho)+\lambda_3(\rho)<2-\varepsilon$ where $\varepsilon$ depends directly on the purity of the global state $\rho$, as described in the Appendix. Reconstructing the complete density matrices, we obtain that the purity of all the experimental global $\mathcal{W}$ states (we are interested in distinguishing genuine three-partite from biseparable entanglement) are higher than 0.87. In this case, the value of $\varepsilon$ is around 0.15, changing the position of the theoretical border $\partial\mathcal{W}$ (as shown if Fig. \ref{FigErr}). To better visualize this, in Fig. \ref{FigErr} we project the 3-dimensional polytopes in a plane. The blue region corresponds to the $\mathcal{W}$ polytope, the $\mathcal{GHZ}$ polytope corresponds to the entire shaded area and the black line corresponds to the biseparable polytope. The mixedness creates the white area, restricting the region of exclusive $\mathcal{GHZ}$ states. One can prove that, as the mixedness increases, the size of this region also increases, making the identification of $\mathcal{GHZ}$ states more difficult using local measurements alone \cite{Walter}. Note that the completely mixed state has the same eigenvalues as the $\mathcal{GHZ}$ state, but in this case the red region collapses to the point $(\lambda_1+\lambda_2, \lambda_3)=(1,1/2)$, making the identification of $\mathcal{GHZ}$ class of states impossible. Nevertheless, this is not the case of our experimental states, implying that we have genuine entanglement of the $\mathcal{GHZ}$ class. Furthermore, the $\mathcal{W}$ states are on the pure state border meaning that these states possess genuine entanglement. The bi-separable states are on the black line that corresponds to the biseparable polytope. Note that the polytope corresponding to the biseparable states also possesses an associated error region related to the impurity of the states that is represented by the dashed line marked with $\partial\varepsilon_{\mathcal{BS}}$ in Fig. \ref{FigWit}. 

\subsection{Four qubits analysis }

In this section we study the entanglement polytopes for four-qubit states. The experimental setup to create and measure inequivalent states of four qubits is similar to the one shown in Fig. \ref{fig:setup}. The central difference is that we introduce another nested interferometer in the path of photon B. A detailed explanation of this setup can be found in the Appendix.  In this case, there are 13 entanglement classes, 7 of which contain genuine four-partite entanglement (being therefore full-dimensional). We have prepared representatives of the states shown in Table \ref{table:qubits}  employing the notation of Ref. \cite{verstraete01} which is also defined in the Appendix. Unfortunately, we cannot draw the complete polytope, which is the convex hull of 12 vertices \cite{Walter,Briand}: ($i$) the vertex corresponding the the product state $(1,1,1,1)$; ($ii$) 6 vertices corresponding to two-partite entangled states which, apart from permutations, can be represented by $(0.5,0.5,1,1)$; ($iii$) 4 vertices of the genuine three-partite $\mathcal{GHZ}$ entanglement, which are the permutations of $(0.5,0.5,0.5,1)$, and the ($iv$) 4 vertices of the genuine three-partite $\mathcal{W}$ entanglement, which are the permutations of $(0.66,0.66,0.66,1)$. $(v)$ An image of the four-partite $\mathcal{GHZ}$ entangled state, the vertex $(0.5,0.5,0.5,0.5)$. Note that this last vertex does not imply that the state is genuinely four-partite entangled. To witness this kind of entanglement we must make sure that the given state does not belong to any other lower-dimensional polytope corresponding to bi-separable states.
\begin{table}[ht]
\centering
\begin{tabular}{|c|c|}
\hline
Family & Local eigenvalues of the prepared state \\
[0.5ex]
\hline
$G_{abcd}$ & $(0.532(6), 0.521(9), 0.524(6), 0.542(8))$ \\
$L_{abc_{2}}$ &  $(0.9967(6), 0.9986(5),0.9934(7), 0.9905(8))$\\
$L_{a_{2}b_{2}}$ & $(0.9922(8), 0.961(1), 0.551(3), 0.552(4))$ \\
$L_{ab_{3}}$ & $(0.696(4),0.805(3), 0.757(4), 0.731(5))$ \\
$L_{a_{2}0_{3\oplus 1}}$ & $(0.682(3), 0.970(1),0.645(3), 0.689(3))$ \\
$L_{0_{3\oplus \bar{1}}0_{{3\oplus 1}}}$ & $(0.594(3), 0.943(1), 0.572(3), 0.533(5))$ \\
[1ex]
\hline
\end{tabular}
\caption{Four-qubit entanglement classes --- The first column specifies the distinct families while the second one accounts for the local maximum eigenvalues of the reduced density matrices using the notation of Ref. \cite{verstraete01}. Some of these families present an infinite number of classes. Therefore, we are referring here to just one of the classes in the family.}
\label{table:qubits}
\end{table}

As we can see from Table \ref{table:qubits}, we can identify the fully separable states (family $L_{abc_{2}}$), the bipartite entangled state (family $L_{a_{2}b_{2}}$). We also have two states that present some form of three-partite genuine entanglement (families $L_{a_{2}0_{3\oplus 1}}$ and $L_{0_{3\oplus \bar{1}}0_{{3\oplus 1}}}$). However, to say that theses states belong to the $\mathcal{W}$ or $\mathcal{GHZ}$ classes, we need further analysis. For this case, we just need to check to which three dimensional polytope each state belongs. For instance for case of the family $L_{0_{3\oplus \bar{1}}0_{{3\oplus 1}}}$ we have
\begin{equation}
\lambda_{1} + \lambda_{2} + \lambda_{3} = 1.699(7) < 1.74 ,
\end{equation}
which tells us that the state contains a genuine three-partite $\mathcal{GHZ}$ kind of entanglement while being separable in the other partition. The value 1.74 was obtained by subtracting from the boundary 2 the associated value of $\varepsilon$, equal to 0.26 in this case. The other state (family $L_{a_{2}0_{3\oplus 1}}$) belongs to the border of the $\mathcal{W}$ and $\mathcal{GHZ}$ classes. Therefore, we can safely say that it presents genuine three-partite entanglement, but we cannot tell apart the class. Through complete quantum state tomography we see that it belongs to the $\mathcal{W}$ class. 

There is a four-dimensional polyhedron (in completely analogy with the three-qubit case) given by the inequality
\begin{equation}
\lambda_{1} + \lambda_{2} + \lambda_{3} + \lambda_{4} < 3,
\end{equation}
that determines the four-partite $\mathcal{GHZ}$ states. Considering the state $G_{abcd}$ that we prepared, we have
\begin{equation}
\lambda_{1} + \lambda_{2} + \lambda_{3} + \lambda_{4} = 2.12(1) < 2.5,
\end{equation}
where, again, the bound was computed using the purity of the state. Therefore, we conclude that the global state has four-partite entanglement $\mathcal{GHZ}$ kind. Regarding our last state (family $L_{ab_{3}}$) we have that (once again in analogy with the three-qubit case)
\begin{equation}
\lambda_{1} + \lambda_{2} + \lambda_{3} + \lambda_{4} = 2.989(8) > 2.5.
\end{equation}
So, our state is on the border of the four-partite $\mathcal{W}$ and $\mathcal{GHZ}$ entanglement classes. Although we know through the witness criteria that this state contains genuine multipartite entanglement, we cannot say what kind of entanglement that may be. By means of the complete quantum state tomography, we verified that we have indeed a four-partite entangled $\mathcal{W}$ state.

\section{Efficiency}

Let us call the method introduced by Ref. \cite{Walter} and experimentally investigated here, the Local Polytope Method (LPM). Even though it does not allow determination of all types of multipartite entanglement, it provides useful information about the entanglement class to which the state may belong. Furthermore, in the case of almost pure states in which it is applicable, it can present some important practical advantages when compared to other entanglement characterization procedures. Its applicability can be checked by making only local measurements \cite{Buhrman, Alves} and obtaining a lower bound for the purity of the multipartite state. Entanglement characterization methods can be divided into two main types:  those that are tomographic in nature and those that are witnesses, providing some limited information about the entanglement in the state.  
\par
Let us first compare the LPM with the tomographic methods. In this case, the practical benefit of the LPM \cite{Walter} is the reduced number of measurements due to the fact that correlations are not measured. In this regard, the LPM requires only independent local measurements, used for determination of the local eigenvalues. For standard tomography \cite{james01}, the number of independent local tomographic measurements on $N$ qubits is $M_{LPM} = 4 N$. This represents an exponential gain when compared to full quantum state tomography (FQST) of a $N$ qubit system, which requires $M_{FQST} = 4^N$. In the case of pure states,  there are more efficient methods, such as compressed sensing and variational techniques that can be used for state tomography (CSQST) \cite{Gross,Maciel}. For these methods, the number of measurements required for $N$ qubits is of the order of $r N^2 2^N$, where $r$ is the rank of the density matrix. Thus, for pure states the number of measurements is $M_{CSQST} \approx N^2 2^N$.  In comparison, the LPM is still exponentially more efficient. A second advantage is that in there is also an exponential gain in terms of the measurement statistics. For example, let us suppose that a source emits $N$ entangled particles with rate $R$, and are detected by $N$ different devices. Due to losses and non unitary efficiency, each particle is detected with an efficiency $\eta$ with $0<\eta<1$. Then, the total $N$-partite count rate for each measurement in the full or compressed-sensing tomographic tecniques is $C_{N} = \eta^N R$. For local measurements, the local count rate used for local tomographic reconstruction of each local density matrix is $C_{local}=\eta R$. This represents an exponential increase in the measurement statistics for $\eta < 1$. For example, for the reasonable value of $\eta=1/4$ and the focusing on the special case $N=4$, the local method is $4^3=64$ times more efficient in terms of registered events. Of course, the tomographic techniques can provide all the information about the density matrix. Still, if the task is to characterize multi-partite entanglement of almost pure states, the LPM could provide a considerable decrease in the number of measurements required.       
\par
Entanglement witnesses are more similar to the LPM, since they typically require less measurements and return only a limited amount of information about the state.  In this regard, the LPM is a multipartite entanglement witness for quasi-pure $N$ qubit states. In fact, as far as we know it is the only witness that requires only local independent measurements. 

When compared to entanglement witnesses, the LPM is interesting in that it may require less measurements, and may be more efficient when lossy detection systems are used.  That is, if the single-qubit detection efficiency is $\eta$, then measuring $m$-party correlations brings a reduction in efficiency of $\eta^m$.  Even if the correlations are measured among fewer systems than available $(m < N)$, such as in Refs. \cite{Brunner, Tura}, the LPM may be advantageous, depending on the number of measurements required for the entanglement witness. That is, the minimum size correlator is $m=2$, and $\eta^2 < \eta$ if $\eta<1$.   Of course, entanglement witnesses may be applicable even in the case of mixed states, whereas the LPM is not. On the other hand, correlation measurements necessarily require the communication of the measurement results of each run, so that the correlation functions can be calculated. All measurements in the LPM are independent.  
\par
Let us quantify this comparison a little further.  We can define a resource ÒoverheadÓ $\mathcal{O}$ as 
\begin{equation}
\mathcal{O}=\frac{\mathrm{number \, of\, measurements}}{\mathrm{efficiency\, of \,measurements}}.
\end{equation}
 For the LPM, we have $O_{LPM}=4N/\eta$.  Let us compare with two types of genuine entanglement witnesses for multipartite systems. To detect genuine multipartite entanglement, one must test the correlations between all the qubits. This can be done in a number of ways. We will choose two extreme cases of witnesses using local measurements: (a) those requiring as few as two correlation measurements on all $N$ qubits (see several examples in Ref. \cite{Brunner}), and (b) those requiring only pairwise correlation measurements on all pairs of systems, giving $N(N-1)$ pairwise measurements in total, such as in Refs. \cite{Brunner,Tura, Klockl}. We see that if applicable, the LPM is already advantageous compared to type (b), since the latter requires about $N^2$ measurements. For entanglement witness type (a), we have overhead $\mathcal{O}(a) =  2 \eta^{-N}$ and for type (b) we have $\mathcal{O}(b)=N(N-1)\eta^{-2}$. Comparing these overheads, one can find a critical detection efficiency for which the LPM is advantageous.  For example, for $N=4$ qubits, the LPM requires less overhead than witness type (a) when  $\eta < 1/2$, and less than witness type (b) when $\eta < 3/4$.  For $N=8$, the LPM is more efficient than type (a) when $\eta  < 0.67$ and always more efficient than type (b).   In fact, it has less overhead than type (b) when $N > 5$.  
 
Performing correlation measurements do not require additional quantum resources as compared to the local measurements.  However, they also require a critical classical resource, which is the synchronization of the measurement bases. For instance, for the polarization of light, one needs to calibrate common vertical and horizontal axes for all parties, and it is impossible to obtain a perfect calibration (it is an asymptotic limit).  Effort has been made to overcome this difficulty, using additional quantum resource in the form of an expanded Hilbert space to encode alignment free qubits \cite{Aolita}. In the multipartite case, we note that in some cases entanglement can be lower bounded  using reference-frame independent correlation tensor norms \cite{Klockl}.

\section{Conclusion}

In conclusion, we have generated and analyzed several types of three and four-partite states using local tomography and a purity bound. We used a recently introduced tool considering incomplete information to characterize these states according to an hierarchy of entanglement classes. We show that it is possible to determine genuine three and four-partite entanglement with this method even in real laboratory conditions, in the presence of small levels of noise.

Even though this scheme does not allow determination of all kinds of multipartite entanglement, it has the great advantage of providing an exponential reduction in the number of measurements required in comparison to full tomographic reconstruction.  Moreover, it also provides an exponential gain in terms of measurement statistics, when measurements are performed with detectors with less than 100\% efficiency.  Since it is based on local tomography alone, it also requires no common reference frame between users.    

We illustrate the usefulness of the method with photons, and we use two degrees of freedom of the same photon to produce different types of multi-partite entanglement. However, the speed up obtained in the identification of multipartite entanglement do not depend on the physical system neither on the degree of freedom used. Therefore, the results obtained here are immediately extended to other systems like ions, superconducting qubits, and atoms, for instance.

{\it Acknowledgements:} Financial support was provided by Brazilian agencies CNPq, CAPES, FAPERJ, and the Instituto Nacional de Ci\^{e}ncia e Tecnologia de Informa\c{c}\~{a}o Qu\^{a}ntica. We thank David Gross, Michael Walter and Matthias Christandl for insightful comments on the first version of the manuscript and Renata Montenegro for helping with the figures. LCC greatly appreciates the warm hospitality of the Universidade Federal do Rio de Janeiro in several visits during the development of this project.

\appendix
\setcounter{equation}{0}

\section{Entanglement classes}
In this appendix, for completeness, we briefly review the definition of equivalence classes of entanglement as well as of the entanglement polytopes and the related witness. Our intention here is not to be exhaustive and we refer the reader to Ref. \cite{durr00,verstraete01,Walter} for a deeper treatment of the subject. 

\emph{Classification of entanglement under SLOCC} --- Two pure density matrices are said to belong to the same class (i.e., they posses the same kind of entanglement) if they can be obtained from each other through invertible local operations and classical communications with a finite probability of success (invertible SLOCC). This sort of classification naturally divides the space of (pure) states in different \emph{equivalence classes}. 

In the case of three qubits, this classification leads to six distinct classes, the two genuinely three-partite entangled $\mathcal{GHZ}$ and $\mathcal{W}$ classes, the three biseparable $\mathcal{BS}$ classes (entangled in the partition AB and separable in C, AB-C, and the analogously defined AC-B and BC-A), and the fully separable one. See the main text for the representatives of all the classes. To see that all of these classes are inequivalent we just need to remember two facts. First, the minimum product decomposition of the $\mathcal{GHZ}$ and $\mathcal{W}$ states are two and three, respectively, which implies that there is no SLOCC protocol to convert one into the other \cite{durr00}. Secondly, as we are dealing with pure global states, the ranks of the reduced density matrices are different in each class (see Table \ref{3qubits}) and, as invertible SLOCC cannot change the ranks of these matrices, we readily see that all the classes are also inequivalent. Note that, if we include non-invertible SLOCC (i.e., at least one of the local operators must have rank one), it is possible to move from a higher class to a lower one, which defines the following hierarchy among the classes \cite{durr00}
\begin{equation*}
\Delta_{\mathcal{S}} \subseteq \Delta_{\mathcal{B}\mathcal{S}} \subseteq \Delta_{\mathcal{W}} \subseteq \Delta_{\mathcal{G}\mathcal{H}\mathcal{Z}}.
\end{equation*}

\begin{table}[ht]
\centering
\begin{tabular}{|c|c|}
\hline
Class & Rank \\
[0.5ex]
\hline
$\mathcal{GHZ}$ & (2,2,2) \\
$\mathcal{W}$ & (2,2,2) \\
\hspace{0.2cm} $\mathcal{AB-C}$ \hspace{0.2cm} & \hspace{0.2cm} (2,2,1) \hspace{0.2cm} \\
$\mathcal{AC-B}$ & (2,1,2) \\
$\mathcal{BC-A}$ & (1,2,2)\\
$\mathcal{S}$ & (1,1,1) \\
[1ex]
\hline
\end{tabular}
\label{3qubits}
\caption{Three qubit entanglement classes --- The first column specifies the distinct classes while the second one accounts for the rank of the reduced density matrices using the notation ($\mbox{rank}(\rho_{A}),\mbox{rank}(\rho_{B}),\mbox{rank}(\rho_{C})$).}
\end{table}

The case of four qubits is much more complicated, presenting nine (up to permutations) equivalence \emph{families} under SLOCC, whose representatives are \cite{verstraete01} 
\begin{widetext}
\begin{equation*}
G_{abcd} = \frac{a+b}{2}\left(|0000\rangle + |1111\rangle\right) + \frac{a-d}{2}\left(|0011\rangle + |1100\rangle\right) + \frac{b+c}{2}\left(|0101\rangle + |1010\rangle\right) + \frac{b-c}{2}\left(|0110\rangle + |1001\rangle\right), 
\end{equation*}
\begin{equation*}
L_{abc_{2}} = \frac{a+b}{2}\left(|0000\rangle + |1111\rangle\right) + \frac{a-b}{2}\left(|0011\rangle + |1100\rangle\right) + c\left(|0101\rangle + |1010\rangle\right) + |0110\rangle ,
\end{equation*}
\begin{equation*}
L_{a_{2}b_{2}} = a\left(|0000\rangle + |1111\rangle\right) + b\left(|0101\rangle + |1010\rangle\right) + |0110\rangle + |0011\rangle ,
\end{equation*}
\begin{equation*}
L_{ab_{3}} = a\left(|0000\rangle + |1111\rangle\right) + \frac{a+b}{2}\left(|0101\rangle + |1010\rangle\right) + \frac{a-b}{2}\left(|0110\rangle + |1001\rangle\right) + + \frac{i}{\sqrt{2}}\left(|0001\rangle + |0010\rangle + |0111\rangle + |1011\rangle\right),
\end{equation*}
\begin{equation*}
L_{a_{4}} = a\left(|0000\rangle + |0101\rangle + |1010\rangle + |1111\rangle\right) + i|0001\rangle + |0110\rangle - i|1011\rangle ,
\end{equation*}
\begin{equation*}
L_{a_{2}0_{3\oplus 1}} = a\left(|0000\rangle + |1111\rangle\right) + |0011\rangle + |0101\rangle + |0110\rangle , 
\end{equation*}
\begin{equation*}
L_{0_{5\oplus 3}} = |0000\rangle + |0101\rangle + |1000\rangle + |1110\rangle , 
\end{equation*}
\begin{equation*}
L_{0_{7\oplus \bar{1}}} = |0000\rangle + |1011\rangle + |1100\rangle + |1110\rangle , 
\end{equation*}
\begin{equation*}
L_{0_{3\oplus \bar{1}}0_{{3\oplus 1}}} = a\left(|0000\rangle + |1111\rangle\right) + |0011\rangle + |0101\rangle + |0110\rangle , 
\end{equation*}
\end{widetext}
where we have employed the notation of Ref. \cite{verstraete01}, in which the sub-indexes of each family are related to the Jordam decomposition of the state and the continuous complex parameters $a$, $b$, $c$ and $d$ (which are the eigenvalues of a complex symmetric matrix) characterizes each family. It is important to observe here that some of these families contain an infinite number of SLOCC classes, none of them accessible in an experiment. This is what happens in general for more than three qubits. In contrast, the developments put forward by Walter \emph{et. al} \cite{Walter} establishes a coarse-grained classification, always presenting a finite number of entanglement classes, which respect a natural hierarchy based on the geometric structure of the entanglement polytopes. Although it is, in general, very difficult to tell apart each one of these classes, it is possible to check for the presence of genuine multipartite entanglement, which is very useful for several applications, especially in quantum information protocols and quantum many-body systems. 

\emph{Entanglement witness} --- SLOCC is equivalent to the existence of local invertible operations, which are represented by matrices of unity determinant, acting on the state space $\mathcal{H}$, naturally constituting a Lie group $G$, the special linear group (see \cite{Bennett1}). Knowing that the orbit $G\cdot\rho$ of an element $\rho \in \mathcal{H}$ relative to the group $G$ is the subset of $\mathcal{H}$ containing the elements to which $\rho$ can be transformed by the action of $G$, the entanglement class containing $\rho$ is then just $G\cdot\rho = \lbrace g\cdot\rho \vert g \in G \rbrace$. In other words, two density operators $\rho$ and $\rho'$ are equivalent if and only if there exist an element $g \in G$ such that $g\cdot\rho = \rho'$. An important property of this definition is that every element of $\mathcal{H}$ belongs to one and only one equivalence class, i.e. given two equivalence classes or they are equal or disjoint. This is a consequence of the fact that two orbits do not overlap.  

As said in the main text, the proposed witness \cite{Walter} is based on the solution of the quantum marginal problem. The set of all possible local eigenvalues compatible with the global state
\begin{equation}
\Lambda = \left(\lambda_{1},...,\lambda_{\mathcal{N}}\right) \nonumber
\end{equation}
forms a convex polytope \cite{Christandl,Klyachko} ($\lambda_{i}$ represents the maximum eigenvalue of partite $i$ reduced density matrix). The authors of Ref. \cite{Walter} noted that, for the cases where the global state is pure, these local eigenvalues contain considerable information about the entanglement of $\rho_{\mathcal{I}}$. They found that the set of possible $\Lambda$ associated with a global state is restricted to a given entanglement class also forms a convex polytope, the so called entanglement polytope. This fact lead the authors to conclude that if $\Lambda_{\rho}$ (the set of eigenvalues of the one-partite reduced density matrix associated with $\rho$) does not belong to a given entanglement polytope $\Delta_{\mathcal{C}}$, then $\rho$ cannot belong to the associated entanglement class $\mathcal{C}$
\begin{equation}
\Lambda_{\rho} \not\in \Delta_{\mathcal{C}} \Rightarrow \rho \not\in \mathcal{C}. \nonumber
\end{equation}

The computation of the entanglement polytopes $\Delta_{\mathcal{C}}$ are based on algebraic geometry and the theory of group representation \cite{Walter}. First, the connection of the SLOCC operations with local invertible operators (Lie group) acting on on the projective Hilbert allowed the computation of the covariants (irreducible subspaces) of such action. Then, by applying the tools from group representation theory, the authors of \cite{Walter} were able to relate these covariants with the eigenvalues of the reduced density matrices of the subsystems thus connecting the entanglement polytopes with the quantum marginal problem. From the fact that the covariants form a finitely generated algebra, it was possible to show that the entanglement polytopes are convex (see \cite{Walter} for the details of the proof and \cite{Wernli} for alternative ways to compute the entanglement polytopes).

\section{Experimental setup for four qubits}

\begin{figure}[tbp]
\centering 
\includegraphics[scale=0.23]{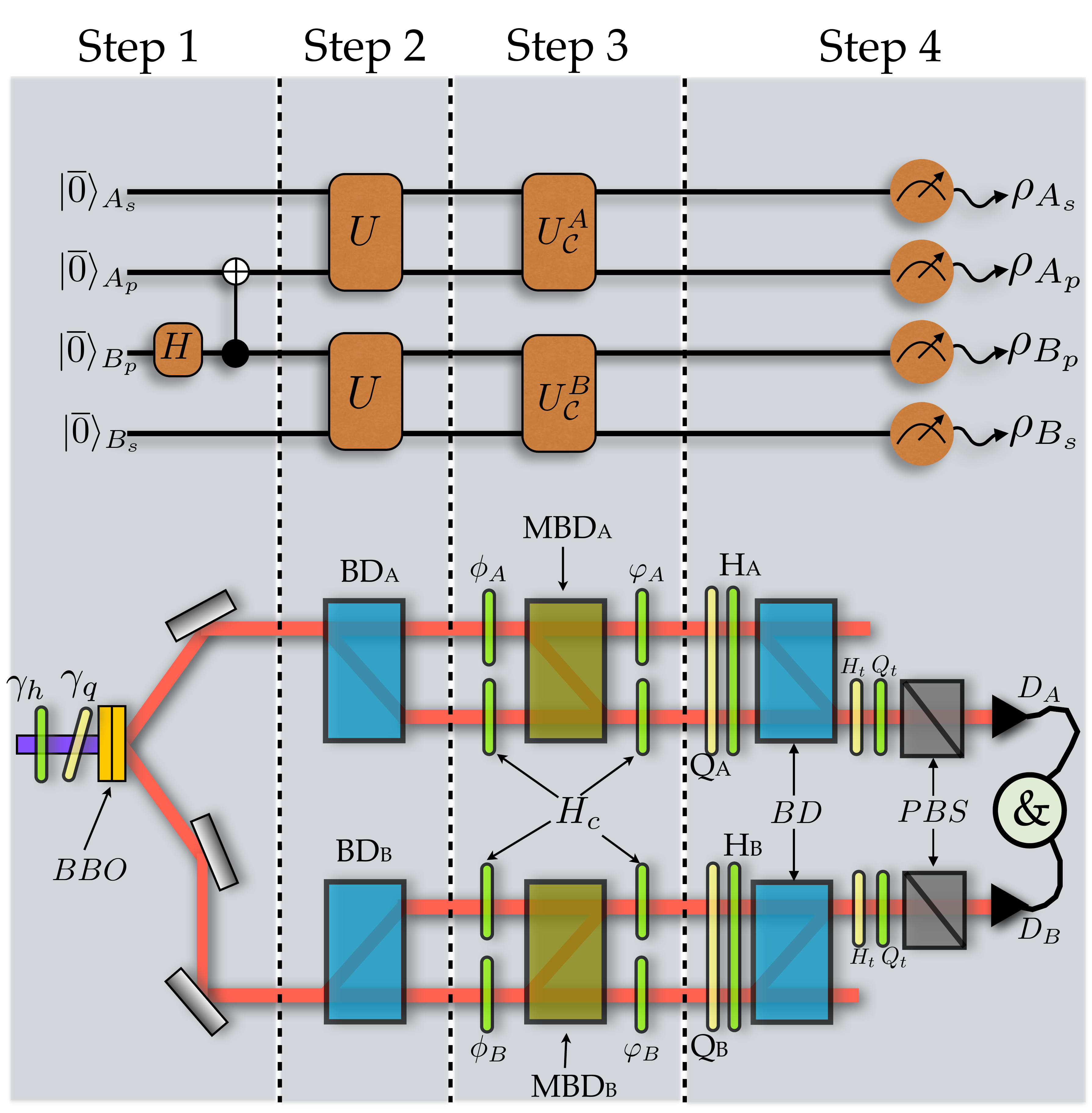}
\caption{Experimental setup for the case of four qubits --- The top panel shows the quantum circuit describing our experiment. In the main text we give a complete description of each step in this circuit as well as of its implementation in the optical scenario, showed in the bottom panel of the figure. The parameters $\alpha$ and $\beta$ appearing in Eq. (\ref{initial}) can be controlled by the half-wave ($\gamma_{h}$) and quarter-wave ($\gamma_{q}$) plates located before the $BBO$ crystals. The $BD$s are beam displacers and $PBS$s are polarized beam splitters. The modified beam-displacer ($MBD$), in contrast with the common $BD$, transmits the horizontal polarization and deflects the vertical one. Physically, it is implemented by putting a $BD$ between two half-wave plates, which are not shown in the figure. The $\theta$ plates are used to ensure identical path lengths in the interferometers. $H_{i}$ and $Q_{i}$ are half- and quarter-wave plates used in the tomographic process, as explained in the text, and the symbol $\&$ represents coincidence counting.}
\label{fig:setup2}
\end{figure}

For the study of the entanglement polytopes in an experimental context for the case of four qubits, we use a  experimental setup similar to the one used in \cite{aguilar14a}. A simplified scenario can be seen in the circuit diagram at the top of Fig. \ref{fig:setup2}. We begin with all the qubits initialized in the state $\ket{\overline{0}}$. In Step 1 we implement a Hadamard ($H$) and a $CNOT$ gates in the qubits $A_p$ and $B_p$, entangling these qubits ($A_s$ and $B_s$ are still separable). In Step 2, two identical unitary operation $U$ are applied, one on qubits $A_p$ and $A_s$ and the other one on $B_p$ and $B_s$. These operations create entanglement between all four qubits. In the third step, we modify the class of entanglement by applying two unitary transformations $U_A$ and $U_B$. In the output, depending on the parameters defining $U_A$ and $U_B$, we can obtain states that are contained in seven of the inequivalent classes of four qubit states \cite{verstraete01}.  In the fourth, and last, step we perform projective measurements and reconstruct the individual reduced matrices $\rho_j$ or the global state $\rho$. We will see later that, with this recipe, we can create and measure states in the class $L_{abc_2}$, $L_{a_2b_2}$, $L_{0_{3\oplus \overline{1}} 0_{3\oplus \overline{1}}}$, $L_{a_2 {0_{3\oplus \overline{1}}}}$, $L_{ab_3}$, and $G_{a bcd}$ \cite{verstraete01}.  

\emph{Step 1} --- The experimental setup can be seen in the bottom of Fig. \ref{fig:setup}. With a laser we pump two cross axes Type I  $\beta$-Barium Borate (BBO) crystals, and creates photons in a state close to \cite{kwiat99}:
\begin{equation}
\ket{\Psi}=(\alpha\ket{0}_{A_p}\ket{0}_{B_p}+\beta\ket{1}_{A_p}\ket{1}_{B_p})\ket{0_s}_{A_s} \ket{0_s}_{B_s}
\label{initial}
\end{equation}
 where $\ket{0}_{A_p}$ ($\ket{1}_{A_p}$) are horizontal(vertical) polarization of the photon $A$. $\ket{0_s}_{B_s}$ represent the spatial degree of freedom of photon $B$. As before, the values of the probability amplitudes $\alpha$ and $\beta$ can be chosen with the half-wave plate $\gamma_h$ and quarter-wave plate $\gamma_q$. Both photons (A and B) are directed to a nested interferometer which implements all operations described in the circuit diagram. 
 
 \emph{Step 2} --- The  two identical unitary operations are applied by the beam displacers $BD_A$ and $BD_B$ which transform $\ket{0}_{i_p} \ket{0}_{i_s}\rightarrow\ket{0}_{i_p} \ket{0}_{i_s}$ and $\ket{1}_{i_p} \ket{0}_{i_s}\rightarrow\ket{1}_{i_p} \ket{1}_{i_s}$, where $i$ can be either $A$ or $B$. After the $BD$s, the state of the photons can be written as
\begin{equation}
\ket{\Psi}_{A_pB_pA_sB_s}=\alpha\ket{0}\ket{0}\ket{0}\ket{0}  +\beta\ket{1}\ket{1}\ket{1}\ket{1}.
\label{GHZ}
\end{equation} 
 We can see that when choosing $\alpha=\beta=1/\sqrt{2}$, the state above is a Greenberg-Horne-Zeilinger (GHZ) state \cite{ghz93}, which is genuinely entangled in all the qubits.  
 
\emph{Step 3} ---We now implement the unitary operations $U_A$ and $U_B$ using a set of half-wave plates (HWP) and a Modified-Beam-Displacer $(MBD_i)$. The transformation for each photon is written in Eq. (\ref{eq:unitary}).  As before, the HWP  $H_{c}$ are used to ensure identical path lengths in the interferometers. As we have two interferometers now, the set of angles is extended to $\mathcal{M}=[\gamma_h, \varphi_1, \phi_1, \varphi_2, \phi_2]$ where different classes of entanglement are obtained for different values of these angles.  

Let us now analyse which states we can create for the different choices of the set $\mathcal{M}$. Suppose that the pump laser is horizontally polarized ($\varphi_h=0$), which is parallel with the axis of one of the BBO crystals. By choosing   $\mathcal{M}=[0,0,0,0,0,0]$ the states are completely separable belonging to the $L_{abc_2}$ class. For the case  $\mathcal{M}=[\pi/4,0,0,0,0,0]$, we obtain a Bell state in the polarization and separable states in the spatial degree of freedom, as given in Eq (\ref{initial}). After $BD_A$ and $BD_B$ the photons are in $GHZ$ state. Since no rotation is applied in the following steps, all the photons exit the first interferometer in the spatial mode $0_{t_i}$. Since there is a coherent superposition of $0_{s_i}$ and $1_{s_i}$ at $MD_i$, the state is still a Bell state in the polarization and separable in the spatial degrees of freedom at output of $U_i$. This state is part of the $L_{a_2b_2}$ class of entanglement. For  $\mathcal{M}=[\pi/4, \pi/4, 0, \pi/4, 0]$, as before, a Bell state is created in the polarization in the step 1. Since $\phi_A=\pi/4$ all the photons in mode $1_{s_A}$ go out the interferometer in the mode $1_{t_A}$. Note that $\varphi_A$ also is equals to $\pi/4$ rotating the polarization of the photons in this mode to $\ket{1}_{A_p}$, see Eq. (\ref{eq:unitary}). Since $\phi_B=0$, all the photons $B$ are coherently combined in the mode $0_{t_B}$. In this case, the photons are in a $\ket{GHZ}_{A_p p_B t_A} \ket{0}_{t_B}$ at the detection step. This states belongs to the family  $L_{{0_{3\oplus \overline{1}}} {0_{3\oplus \overline{1}}}}$. For $\mathcal{M}=[\pi/3, \pi/8, 0, 0, 0]$, $\alpha=\sqrt{1/3}$ and $\beta=\sqrt{2/3}$ at the initial step. Since $\phi_A$ is $\pi/8$, part of the photons of mode $1_{s_A}$ goes out the interferometer in the mode $0_{t_A}$ and the other part in $1_{t_A}$. Since no rotations are implemented in the latter modes, the states at the detection step is $\ket{W}_{A_p p_B t_A} \ket{0}_{t_B}$ where $\ket{W}=\sqrt{1/3}(\ket{001}+\ket{010}+\ket{100})$ defined in \cite{durr00}. This state is part of the family $L_{a_2 {0_{3\oplus \overline{1}}}}$. Following the same procedure, we can demonstrate that for $M=[\pi/2, \pi/8, \pi/8, 0, 0]$, the state is $\ket{W_{A_p p_B t_A t_B}}=1/2(\ket{0001}+\ket{0010}+\ket{0100}+\ket{1000})$  which corresponds to the family $L_{ab_3}$ . For $\mathcal{M}=[\pi/4, \pi/4, \pi/4, \pi/4, \pi/4]$ is a GHZ state as the one defined in Eq. (\ref{GHZ}). This state belongs to the family $G_{abcd}$. Finally, for $\mathcal{M}=[\pi/4, \pi/4, \pi/4, \pi/4, 0]$ the emergent state is part of the family $L_{0_{5\oplus \overline{3}}}$ of Ref. \cite{verstraete01}.

\emph{Step 4} --- After the engineering of different entanglement classes, we made projective measurements in the different degrees of freedom. For this, in every photon we put a set composed of a BD, two QWP, two HWP and a polarized beamsplitter (PBS). We me make tomographic measurement of the spatial degree of freedom of both photons using the $H_{t_i}$ and $Q_{t_i}$. After the projection in the polarization and spatial mode, the photons are detected in Det$_A$ and Det$_B$.

\section{Error analysis}

\emph{State preparation} --- To compute the errors bars of the eigenvalues and of the purity we assume that the coincidence counts were distributed accordingly to a Poissonian distribution. We then apply Monte Carlo simulation to obtain a distribution of negativities, taking the standard deviation as the error. We attribute the small error bars to the high number of coincidence counts. Note that the Poissonian count statistics is not the main error concerning the application of the witness. The value of $\varepsilon$ depends on the fidelity of the prepared state relative to a given theoretical pure state. The procedure employed to compute this error is explained below.    

\emph{The value of $\varepsilon$} --- The witness investigated here is based on the assumption that we have a pure global state. If the global state is mixed, than any set of local density matrices would be possible and we could not expect to extract global information of the total state from local information of the parts. However, as commented in the main text, this witness is robust against some levels of noise. Here we explain how we computed the confidence boundary shown in Fig. \ref{FigErr}.

Denoting by $p$ the purity of the prepared state, it was shown in REf. \cite{Walter} that the vectors of the local eigenvalues of a pure state differs from that of the prepared one by at most $N(1-\sqrt{2p-1})$, as long as $p>1/2$ ($N$ is the number of parties). This was proved by analysing the trace norm between both set of eigenvalues. For the case of qubits (in which a variation of the maximum eigenvalue must be accompanied by an opposite variation of the other one), this bound can be further improved resulting in the boundary shown in the main text
\begin{equation}
\varepsilon = \frac{N}{2}\left(1 - \sqrt{2p-1}\right).
\end{equation}

\begin{figure}[h]
\includegraphics[scale=0.34]{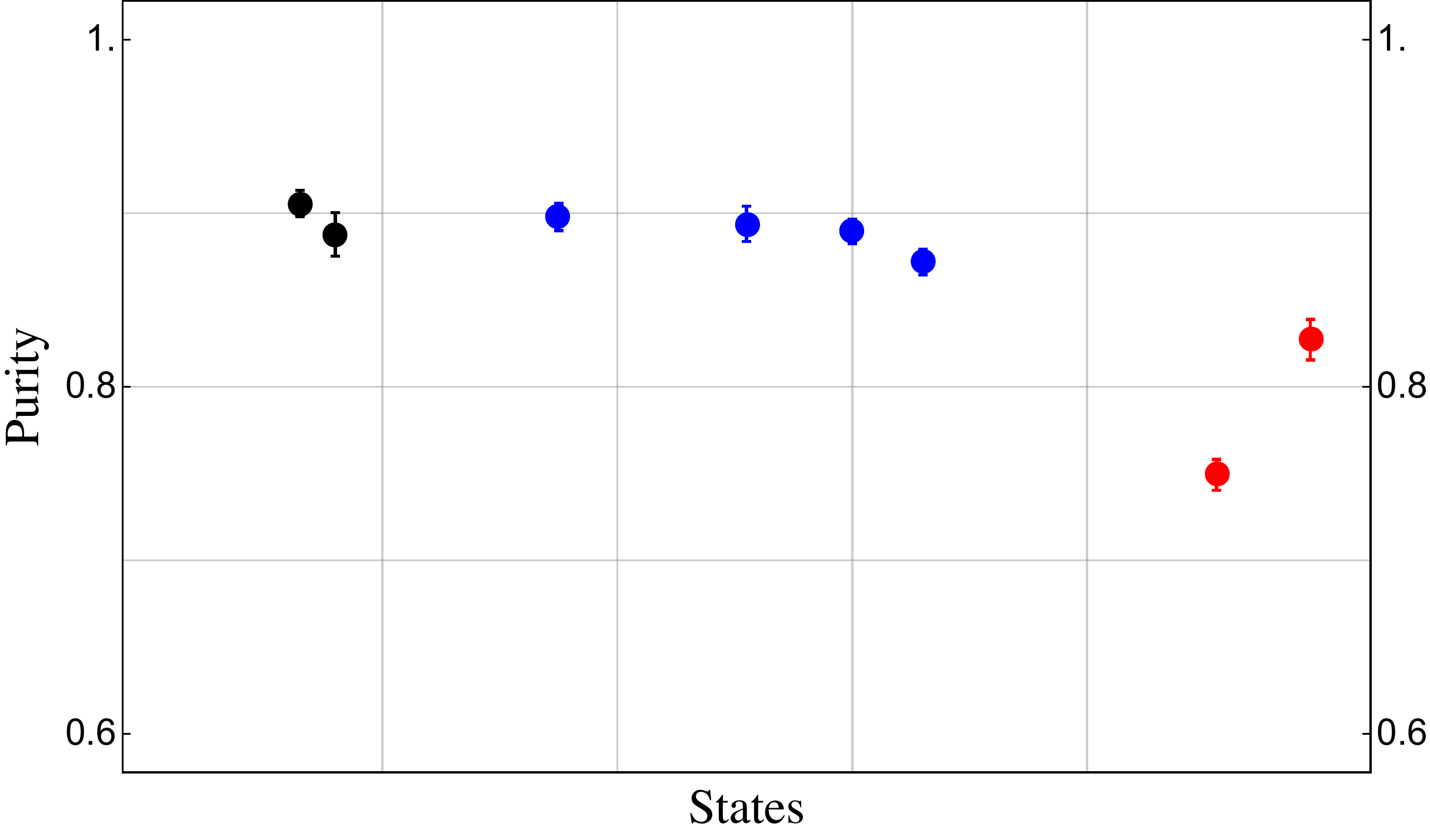}
\caption{Purity of the 3 qubit states --- The blue, red and black dots are the $\mathcal{W}$, $\mathcal{GHZ}$ and $\mathcal{BS}$ states prepared, respectively. Error bars are also shown}
\label{Purity3q}
\end{figure}

\begin{figure}[h]
\includegraphics[scale=0.36]{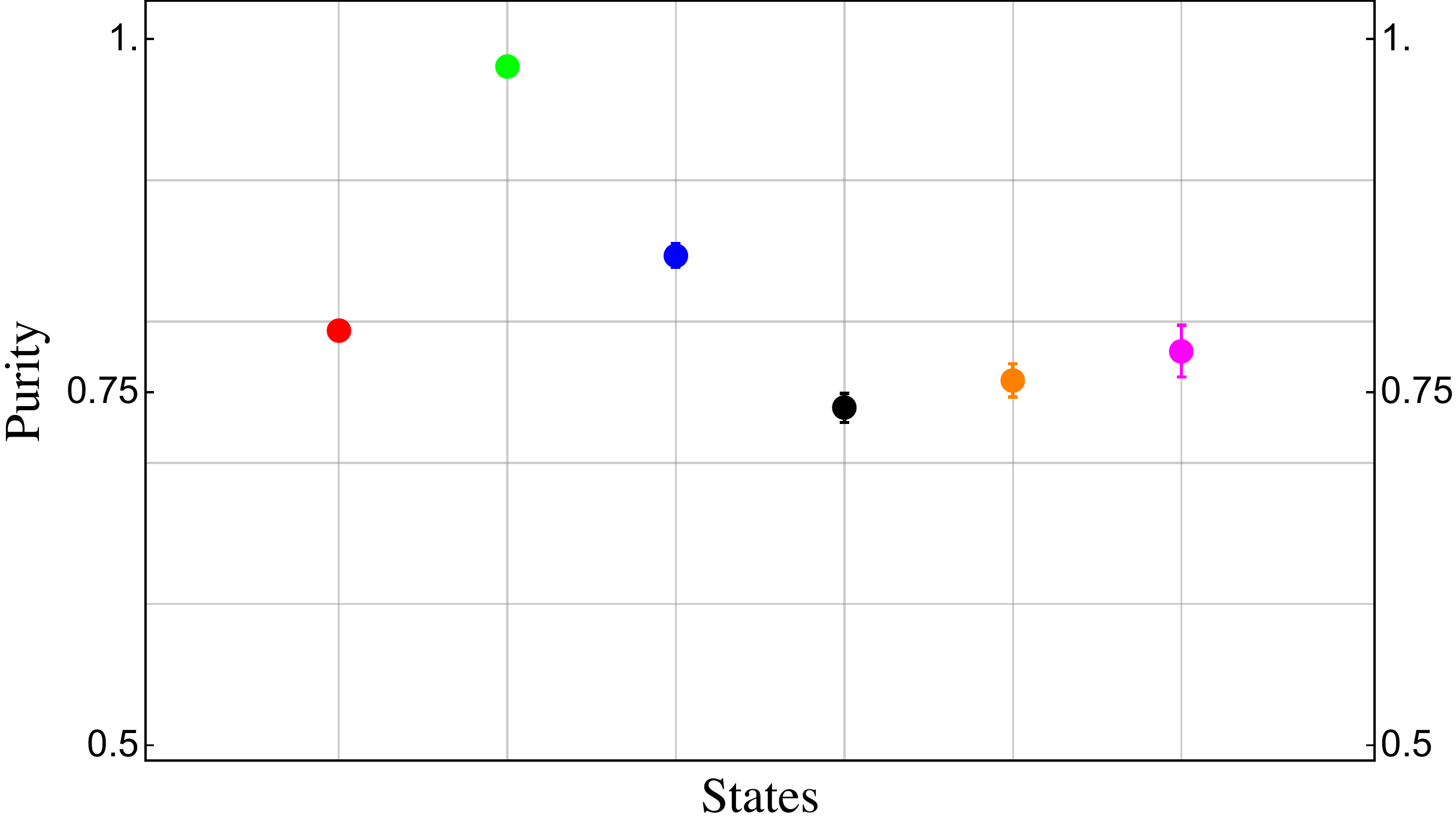}
\caption{Purity of the 4 qubit states --- Representatives of the $L_{a_{2}b_{2}}$ (red), $L_{abc_{2}}$ (green), $L_{0_{3\oplus\bar{1}}0_{3\oplus\bar{1}}}$ (blue), $L_{a_{2}0_{3\oplus\bar{1}}}$ (black), $L_{ab_{3}}$ (orange) and $G_{abcd}$ (magenta) families. The error bars are also shown.}
\label{Purity4q}
\end{figure}

Note that we need to know the value of the purity of the global state. This is a nonlinear function of $\rho$ and therefore cannot be obtained by means of local measurements. However, it is possible to get a lower bound on this quantity if more copies of $\rho$ are available \cite{Buhrman}. Figures \ref{Purity3q} and \ref{Purity4q} show the purity for our three and four prepared states, respectively.  These values are reasonably high, showing that our scheme is suitable not only for the verification of the entanglement witness but also to be employed in situations far beyond the present work.

Remember that we chose the worst value of the purity among all the prepared states to compute the value of $\epsilon$.

\end{document}